\documentclass[twoside,twocolumn,10pt]{article}

\usepackage{wscg}           
\usepackage[backend=biber,
            style=alphabetic,
            maxnames=99,   
            minalphanames=1, 
            giveninits=false 
            ]{biblatex} 
\RequirePackage{ifpdf}
\ifpdf
 \RequirePackage[pdftex]{graphicx}
 \RequirePackage[pdftex]{color}
\else
 \RequirePackage[dvips,draft]{graphicx}
 \RequirePackage[dvips]{color}
\fi

\usepackage{nopageno}       

\usepackage{amsmath,amssymb}
\usepackage{booktabs}
\usepackage{algorithmic}
\usepackage{algorithm}
\usepackage{dblfloatfix}
\usepackage[switch]{lineno}
\usepackage{graphicx}    
\usepackage{subcaption}  

\usepackage[pscoord]{eso-pic}
\usepackage[color=black,opacity=1,angle=0,scale=1]{background}
\newcommand{\placetextbox}[3]{
  \setbox0=\hbox{#3}
  \AddToShipoutPictureFG*{
    \put(\LenToUnit{#1\paperwidth},\LenToUnit{#2\paperheight}){\vtop{{\null}\makebox[0pt][c]{#3}}}%
  }%
}%
\backgroundsetup{
  contents={%
  \placetextbox{0.5}{.95}{Accepted WSCG 2026}
  \placetextbox{0.5}{0.085}{\thepage}%
  }
}

\title{Multi-planar 2D-U-Net Segmentation of 3D-CT Abdominal Organs augmented by Spatial Occurrence Maps}

\author{\hspace{-.5cm}\scriptsize
\parbox{0.25\textwidth}{\centering
Daria Kern\\[1mm]
Glasgow Caledonian University\\
School of Science \\
\& Engineering\\
Glasgow G4 0BA, \\United Kingdom\\[1mm]
daria.kern@gcu.ac.uk
}
\hspace{0.00\textwidth}
\parbox{0.25\textwidth}{\centering
Negar Chabi\\[1mm]
Jade University of Applied Sciences\\
Department of Engineering\\
\& Medical Technology\\
Germany, 26389, Wilhelmshaven, Germany\\[1mm]
negar.chabi@jade-hs.de
}
\hspace{0.00\textwidth}
\parbox{0.25\textwidth}{\centering
Souraj Adhikary\\[1mm]
Jade University of Applied Sciences\\
Department of Engineering\\
\& Medical Technology\\
Germany, 26389, Wilhelmshaven, Germany\\[1mm]
souraj.adhikary@student.jade-hs.de
}
\hspace{0.00\textwidth}
\parbox{0.25\textwidth}{\centering
Andre Mastmeyer\\[1mm]
Jade University of Applied Sciences\\
Department of Engineering\\
\& Medical Technology\\
Germany, 26389, Wilhelmshaven, Germany\\[1mm]
andre.mastmeyer@jade-hs.de
}
}

\usepackage{url}
\makeatletter
\def\Uslash{\mathbin{\mathchar`\/}\@ifnextchar{/}{\kern-.15em}{}}
\g@addto@macro\UrlSpecials{\do \/ {\Uslash}}
\def\Ucolon{\mathbin{\mathchar`:}\@ifnextchar{/}{\kern-.1em}{}}
\g@addto@macro\UrlSpecials{\do : {\Ucolon}}
\makeatother
\renewcommand{\marginpar}[1]{}
\renewcommand{\textcolor}[1]{}

\begin{document}

\twocolumn[{\csname @twocolumnfalse\endcsname

\maketitle  

\begin{abstract}
\noindent
This work proposes a lightweight 2D-U-Net-based framework for segmenting five abdominal organs in large field-of-view 3D CT scans. The method combines coarse-to-fine segmentation, predictions from multiple anatomical planes, and additional fuzzy 3D spatial maps that provide anatomical location cues to improve segmentation accuracy. We combine multi-planar 2D-U-Net models augmented by a spatial occurrence map. The approach involves two main stages. First, the abdominal volume of interest region is detected by traversing the whole scan axially with a 2D-U-Net and determining the x-y-z-minimum and -maximum extents of the 5 abdominal organs of interest. Second, we use spatial occurrence maps to enhance our multi-planar 2D-U-net architecture inside the bounds from the former stage. The method is evaluated on 80 CT scans from various public sources. The results show Dice improvements of about 4\% at maximum compared to the same model trained without spatial occurrence maps.
\end{abstract}

\subsection*{Keywords}
Abdominal CT segmentation, 2D-U-Net, Priors, Spatial Occurrence Map, Multi-Channel-U-Net
\vspace*{1.0\baselineskip}
}]


\section{Introduction} \label{sec:intro}
\textcolor{green}{Accurate abdominal multi-organ segmentation \cite{llopis2026diverse} with field-of-view (FoV) normalization is a key step toward reliable image-guided clinical decision-making. It plays an important role in human diversity-sensible computer-aided diagnosis, surgical navigation, and radiation therapy. However, it remains challenging due to the varying CT scan and organ group fields of view, low soft-tissue contrast, and the close spatial proximity of organs and neighboring visceral structures. A particular challenge is the high variability in the shape and intensity appearance of abdominal structures, as organs do not look exactly the same in every person or scan depending on contrast agents and food or nutrition state. Furthermore, some structures are difficult to distinguish even for experts with regards to generating accurate ground truth segmentation masks. For instance, the pancreas has an irregular shape and often blends by intensity and touches nearby soft tissues or concurring organs of the target group (liver, spleen, right kidney, left kidney, and pancreas).}
\copyrightspace

\textcolor{red}{Convolutional Neural Networks (CNNs) are extensively utilized in medical image segmentation due to their capacity to extract hierarchical features from complex medical imaging datasets.}
\textcolor{red}{They have transformed the field by enabling automated learning of features that help clinicians accurately determine organs, lesions, their sizes, locations and boundaries, and relationships with surrounding tissues.}
\textcolor{red}{CNNs address the unique challenges in medical imaging, such as low resolution and poor contrast, by using convolutional layers to capture spatial hierarchies and pooling layers to reduce dimensionality while preserving key image features.}
\textcolor{red}{This has led to major advances in diagnostic precision and patient care, making CNNs a state-of-the-art technique for tasks such as disease detection and organ segmentation
\cite{bertels2022convolutional,khobragade2024comprehensive, mienye2025deep,gao2025medical}.
} 
The 2D U-Net architecture in Ronneberger et al. \cite{0000-2dunet} and the 3D version by \c{C}i\c{c}ek et al. \cite{0000-3dunet} set decisive milestones. Unlike previously known CNNs, the U-Net concept uses down- and up-samplings sharing skip connections, incorporating higher resolution features in up-sampling steps.

\textcolor{green}{As medical imaging datasets continue to grow, increasingly complex CNN architectures have been developed, some containing hundreds of millions of trainable parameters.}
\textcolor{green}{Instead of increasing model complexity, we have introduced a more lightweight pipeline in \cite{0000-2dvs3dunet} and augment it here. Although the aforementioned challenges remain, coarse organ location can provide a useful anatomical prior, reducing the reliance on image intensity (one gray channel) alone. Abdominal organs occur in relatively predictable regions of the body. We exploit this property by providing the model with additional spatial cues through spatial occurrence maps ($SOM$). This kind of map is provided to an ensemble of 4 2D-U-Nets as an additional input channel, so the network receives both the CT image and prior anatomical location information. The proposed method is computationally efficient, can run on consumer GPUs (e.g. 16 GB VRAM), and has very fast prediction time while improving segmentation results.}

\begin{figure*} [!ht]
   \centering
    \begin{subfigure}[t]{.325\textwidth}
        \centering
        \includegraphics[width=\linewidth]{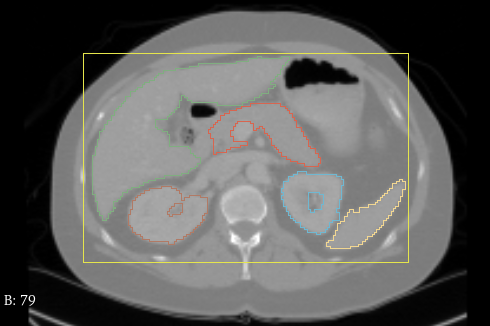}
        \caption{axial slice}
        \label{fig:1sub_a1}
    \end{subfigure}
    \hfill
    \begin{subfigure}[t]{.325\textwidth}
        \centering
        \includegraphics[width=\linewidth]{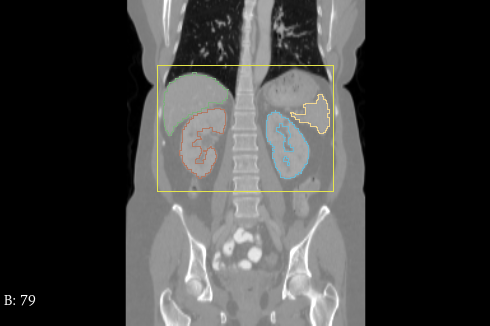}
        \caption{coronal slice}
        \label{fig:1sub_b1}
    \end{subfigure}
    \hfill
    \begin{subfigure}[t]{.325\textwidth}
        \centering
        \includegraphics[width=\linewidth]{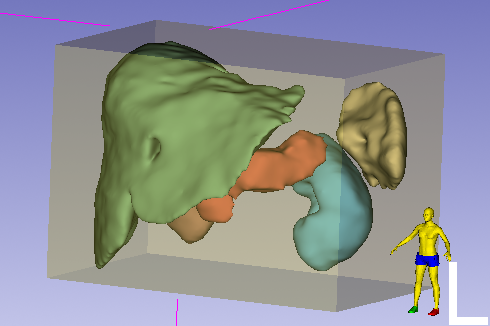}
        \caption{3D overview}
        \label{fig:1sub_c1}
    \end{subfigure}
\vspace{2mm}
    \caption{Summary overview: (a) axial slice and bounding box (yellow rectangle) detected in stage 1. (b) coronal slice and bounding box as well as organ contours: liver=green, spleen=light orange, right kidney=brown, left kidney=blue, pancreas=red. (c) 3D overview of stage 2 organs to be found inside the bounding box of stage 1.}
    \label{fig:overview}
\end{figure*}

\textcolor{blue}{\marginpar{3.6, 3.12}\textbf{New contribution:} In this work on 2D-U-Net segmentation of 5 abdominal organs in 3D CT image data ranging from the upper lungs to the hip, we propose a multi-stage (coarse to fine) and multi-planar ensemble of 2D-U-Nets - with a special focus on augmentation by a fuzzy spatial 3D map designs serving as additional input channel to improve Dice results, see Fig.~\ref{fig:overview}}.

\section{Related Work} \label{sec:related}
In training 3D CNN architectures on full-body CT scans, several hierarchical approaches have emerged that cascade phases from coarse to fine \cite{0000-nnUnet, 0000-hier3d, 0000-CNNbasedHC}. Other works combine 2D and 3D architectures \cite{0000-denseunet, 0000-hybrid2,0000-an2025multi} to hybrid approaches. Yet under some circumstances, handy 2D-U-nets can outperform the theoretically superior 3D U-Nets (more training parameters), which we have already investigated in previous works \cite{0000-kern20212d,0000-kern20223d, 0000-2dvs3dunet}. \textcolor{blue}{\marginpar{2.4} Work on combining attention maps with CNNs started in 2017 \cite{vaswani2023attentionneed}.}\textcolor{blue}{\marginpar{2.4, 3.1} Compared to our own previous work focusing 2D vs. 3D U-Nets in this conference \cite{0000-2dvs3dunet}, we yield improvements of 1-2\% (main organs) and 10\% for pancreas, as well as smaller deviations around means and medians, i.e. more precise results.} \textcolor{blue}{\marginpar{2.3, 2.4, 2.7} Up to date Dice coefficients are always shown in database challenge leader-boards\footnote{e.g.~AMOS~-~\\https://amos22.grand-challenge.org/evaluation/amos-ct-regular-evaluation/leaderboard/}, with e.g. a difference of up to 3\% (liver=0.98) compared to our results in Tab. \ref{tab:dice}. \textcolor{blue}{\marginpar{3.10} In the current literature, whole body segmentation has gained some progress. Peer-reviewed references report recent Dice scores in \cite{he2025vista3d,guo2025towards} 0.96 for e.g. liver and spleen.} Thus, our pipelines' results in Table \ref{tab:dice} are deemed competitive, representative and most importantly here, as solid ground for sound and complete experimentation in our single factor ablation study design.} Hence, we take it from our previous work and continue this line by improving one interesting selected aspect, as focused in the research question.

\section{Research Question} \label{sec:quest}
Could the factor "spatial occurence map channels" augment the segmentation quality of U-Nets, exemplified by a 2D-U-Net pipeline in this study?

\section{Method}\label{sec:method}

\subsection{Overview} \label{sec:over}
Our method (Fig. \ref{fig:overview}) can be summarized in two steps: 
\textcolor{green}{(1) The abdominal area is first localized using a 2D-U-Net that operates on axial slices only and predicts an organ group volume of interest (VOI). We define this area by the maximum extents of the spleen, liver at the top, and kidneys at the bottom.}
(2) A 2nd spatial occurrence map ($SOM$) is generated to encode z-voxel depth and frequency average of organ voxels accumulated from training set labels. This probability map serves as a second input channel for an ensemble of 3 2D-U-Nets for axial, coronal and sagittal sections. A maximum label probability fusion step concludes the pipeline.

\subsection{Preprocessing} \label{sec:pre}
For training and testing, we use 80 CT patient scans and label data from various public sources\footnote{VISCERAL - https://service.tib.eu/ldmservice/dataset/visceral-dataset\\ SLIVER - https://sliver07.grand-challenge.org\\ LITS - https://www.kaggle.com/datasets/andrewmvd/liver-tumor-segmentation\\ AMOS - https://amos22.grand-challenge.org}. \textcolor{blue}{\marginpar{3.13}All scans are resampled to isotropic voxels with size of 2 mm$^3$ (2 mm cubes)}. Each scan's body region is separated from the background by thresholding with -200 HU. Segmentation labels undergo morphological closing, and extraction of the 5 largest connected components as the foreground objects. Then intensity values in the scan data set are normalized by pruning to [0.5, 99.5] percentiles, followed by z-score normalization based on the mean and standard deviation. Finally, intensities are rescaled to [0, 1].

\subsection{Abdominal VOI detection} \label{sec:ard}
\begin{figure*} [!ht]
   \begin{center}
   \begin{tabular}{c} 
   \includegraphics[width=\textwidth]{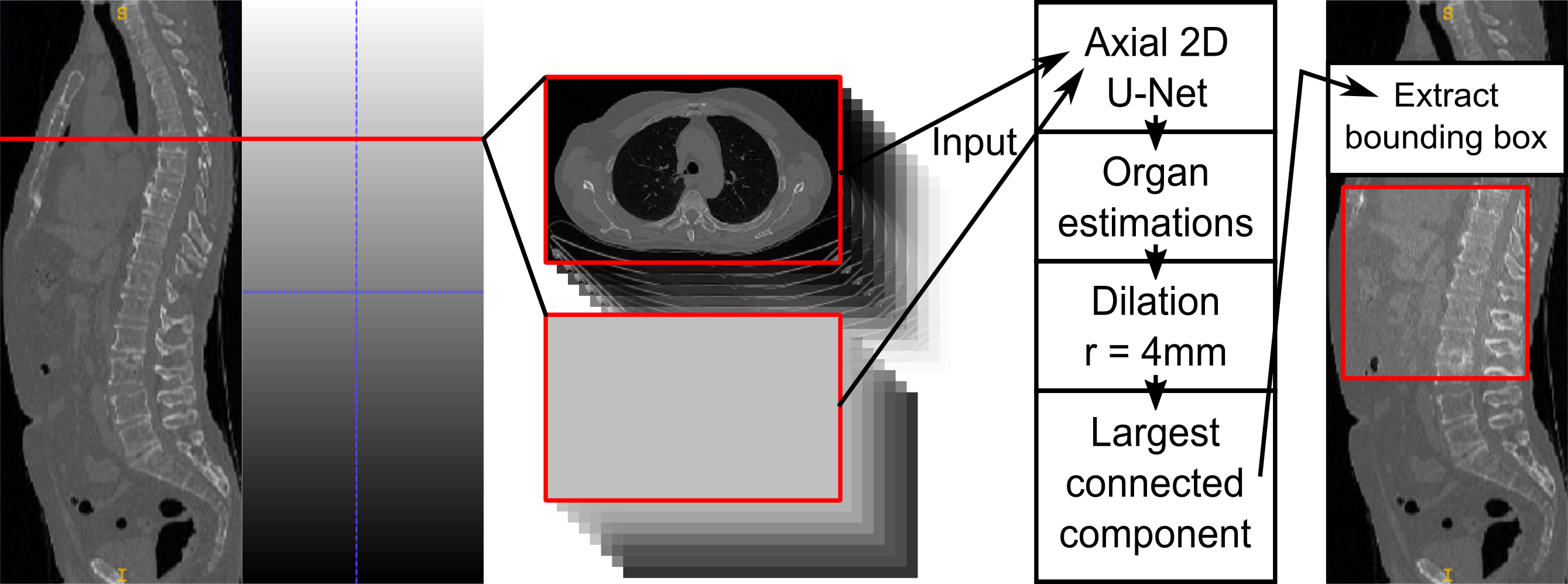}
   \end{tabular}
   \end{center}
   \caption[example] 
   { \label{fig:ard} 
Overview of the steps performed to extract the abdominal region. Left: Sagittal view of patient. Middle: Axial slices and spatial map. Right: VOI detection pipeline and result indication (red box). White (1.0) in the map channel means high probability of organ structures.}
\end{figure*}

\begin{figure*} [!ht]
   \begin{center}
   \begin{tabular}{c} 
   \includegraphics[width=\textwidth]{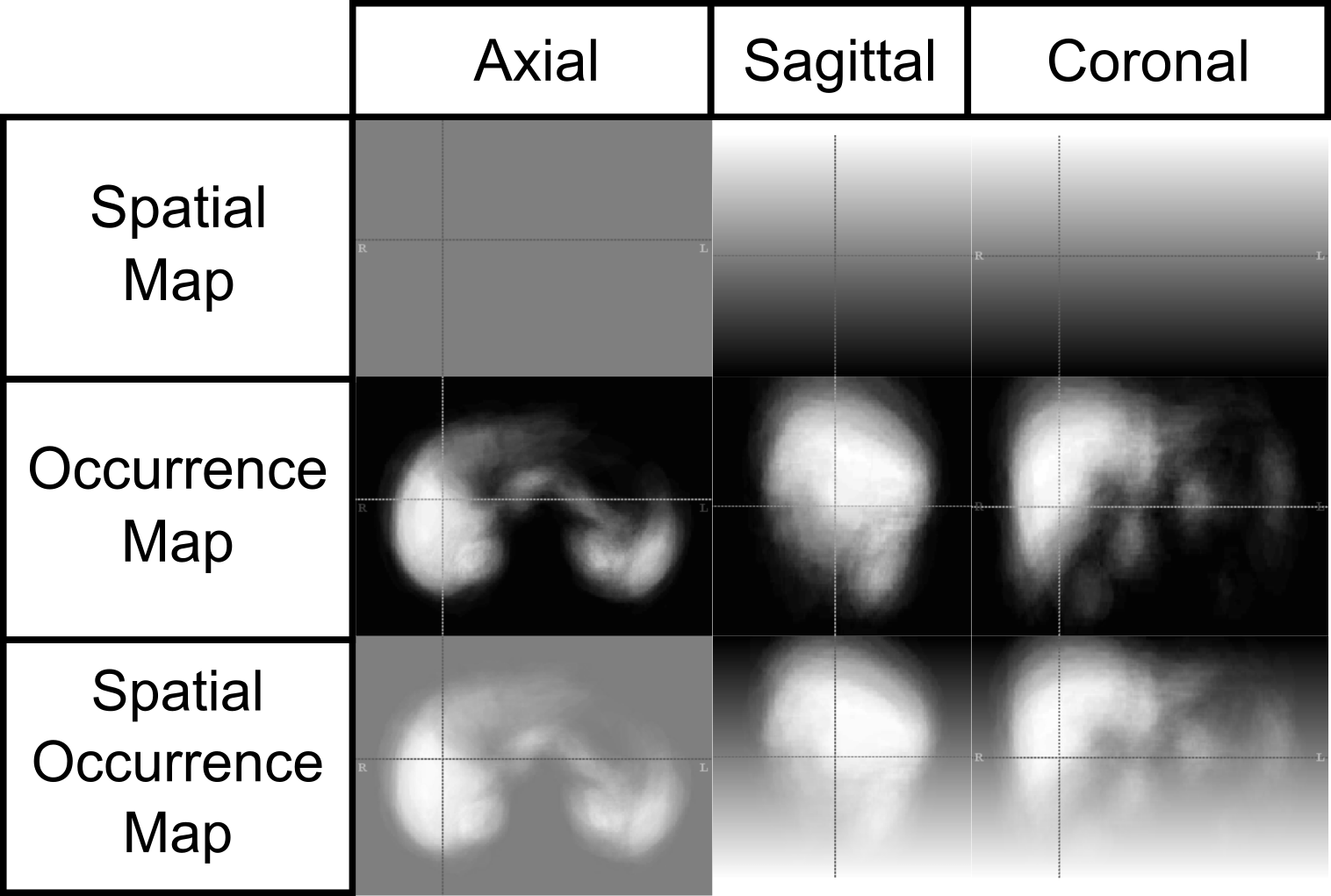}
   \end{tabular}
   \end{center}
   \caption[example]
   { \label{fig:som} 
Overview of fuzzy membership function, spatial map, occurrence map and the combined spatial occurrence map visualized in axial, sagittal and coronal view. White in $SOM$=high probability (1.0) of organ structures.}
\end{figure*}

Since there are major variations in the field of view of the scans, we have to crop the scans to the organ group box
to isolate the abdominal region containing liver, spleen, kidneys and pancreas (Fig. \ref{fig:1sub_a1}, b, c):
First, for each individual scan $S_i$ ($i=1..80$) we generate a fuzzy membership function \textcolor{blue}{(inferior=1 to superior=0,\marginpar{2.5a} Fig. \ref{fig:ard}, left group, 2nd}) represented as a linear list with $n_{i}$ (scan number of slices) elements. Each element represents a heuristic organ structure probability value ranging from 0 to 1:
\begin{equation}
\label{eq:linspace}
    f(n_i) = \textrm{linspace}(0.0, 1.0, n_{i}).
\end{equation}
\textcolor{blue}{\marginpar{2.6a}The motivation for high probabilities on top (superior=1, Fig. \ref{fig:ard}, middle group, 2nd below) of the linear map supports the observation that more candidate voxels of the organ group are in the upper area (liver and spleen mass).}
Second, the 1D linear list is resampled into a 3D fuzzy spatial map  \textcolor{blue}{\marginpar{2.5a}for each individual scan of variable z-axis size.}
Consequently, we define a map $M_i$ as an empty image with the same size as $S_i$ and with constant axial intensity according to $f(n_i)$. Each ($j$-th) axial $M_{i}$ slice \textcolor{blue}{is generated automatically as:}
\begin{equation}
\label{eq:linimagestack}
    M_{i}[j]=f(n_i)[j], \quad j = \{0, ..., n_{i} - 1\}.
\end{equation}
\textcolor{blue}{\marginpar{2.5a}The corresponding image to Eq. \ref{eq:linimagestack} is Fig. \ref{fig:ard}, 2nd group, bottom).}
Thus, for each $M_{i}[j]$, we set all voxels to the corresponding value in the fuzzy membership function $f(n_i)[j]$ (Fig. \ref{fig:som}, top row).
Next, all scans, labels $>0$ (binary) and maps are provided as two input channels
to an axial 2D-U-Net, which is trained over the entire body 
to predict the organ group voxels (liver, spleen, kidneys and pancreas). The architecture consists of five down- and up-sampling steps, followed by a softmax function to calculate the particular organ probabilities. We train the network with a combination of Dice and cross-entropy loss as proposed in \cite{0000-nnUnet}:
\begin{equation}
\label{eq:loss}
    \mathcal{L} = \mathcal{L}_{dice} + \mathcal{L}_{CE}.
\end{equation}
Training takes place for 250 epochs with a batch size of 16 and a learning rate of 0.0001, which we reduce by $50\%$ if the loss does not decrease for 30 epochs. For abdominal region extraction, we morphologically dilate all organ predictions by a radius of $r = 4$ mm, reduce them to the 5 largest connected components (number of target organs, cf. Fig. \ref{fig:ogr}) and find the 3D bounding box.

This model yields VOI-predictions for the whole organ group (Fig. \ref{fig:ard}) to be processed by subsequent models.

\subsection{Multi-planar organ segmentation} \label{sec:multi-planar}
For the next stage, our new approach is to incorporate anatomical probability into the segmentation process by constructing an organ occurrence map and combining it with previous spatial maps $M_{i}$ (Sec. \ref{sec:ard}).
To this end,
we calculate an occurrence map $O$ over all label images in the training set in a normalized scan space.
All label images are first
cropped according to the extracted VOI and thresholded $>0$ to obtain binary masks (1 = organ present, 0 = background). 
We then 
calculate the median slice count $n_{med}$ over the whole dataset obtained from 
the first stage (Section \ref{sec:ard}) and resample all cropped images in $z$ direction to equal size $n_{med}$. After padding the label images in $x$ and $y$ directions to equal size, we initialize the occurrence map $O$ as a zero map of this size. Furthermore, we iterate through all label images, incrementing each voxel of $O$ by 1 at positions corresponding to organ voxels. 
After finishing the accumulation, we divide each voxel of map $O$ by the total number of training images to normalize the map voxels to an organ probability $[0,1]$ (Fig. \ref{fig:som}, middle row).
Then, we combine $O$ with the inverted heuristic map $M_i$ from Sec. \ref{sec:ard} for each scan. We iterate through all training and test scans and 
resample $O$ to fit the 3D size of the scan $S_i$. Thus, we yield a more organ shape-specific occurrence map $O_i$. We then calculate the spatial occurrence map $SOM_i$ by combining 3 congruent components: (1) the scan-specific occurrence map $O_i$, which encodes the typical organ locations derived from the training data, (2) the spatial map $M_i$ representing the abdominal region obtained in the previous stage, and (3) a constant-valued image of ones $[1]_i$:
\begin{equation}
\label{eq:SOM}
    SOM_i = \Big([1]_i - M_i\Big) \odot O_i,
\end{equation}
where $\odot$ denotes voxel-wise maximum operator (Fig. \ref{fig:som}, bottom row). The spatial map $M_i$ is first inverted (Fig. \ref{fig:som}, top row) and then combined with the occurrence map (Fig.~\ref{fig:som}, middle row). The linear map inversion $[1]_i - M_i$  is motivated by the fact that the incorporated occurence map $O_i$ already (middle row, see Fig. \ref{fig:som}) already reflects the organ masses of liver and spleen and we want - \textcolor{blue}{\marginpar{2.6b}in contrast} - now to give more attention to the lower abdominal part where smaller organs such as kidneys and pancreas are seated.

Lastly, we train an ensemble of 2D-U-Net models in axial, coronal and sagittal planes using the scan's, labels and $SOM$s. Each planar model has two input channels for the oriented scan and $SOM$ slices. Since the extracted VOIs from the first stage are clearly smaller vs. the full scan size, our model architecture here consists of only four down- and up-sampling steps. Due to faster convergence here, we train each model only for 100 epochs. All other parameters are the same as in the VOI localization section \ref{sec:ard}. We fuse the results of all 3 oriented 2D-U-Net ensemble models in 3D space by probability maximum for each label. To clean for sparse voxels and isolated islands, the fused 3D label map is reduced to the 5 largest connected components. The auxiliary frameworks used are TensorFlow/Keras 2.20, VTK 9.1 for bounding box calculations and SimpleITK 2.1 for basic image processing and the Dice metric (DSC). Python version for scripting is 3.10 with Ubuntu 22.04 LTS.

\subsection{Evaluation and metrics} \label{sec:eval}
\begin{figure*} [!ht]
   \begin{center}
   \begin{tabular}{c} 
   \includegraphics[width=\textwidth]{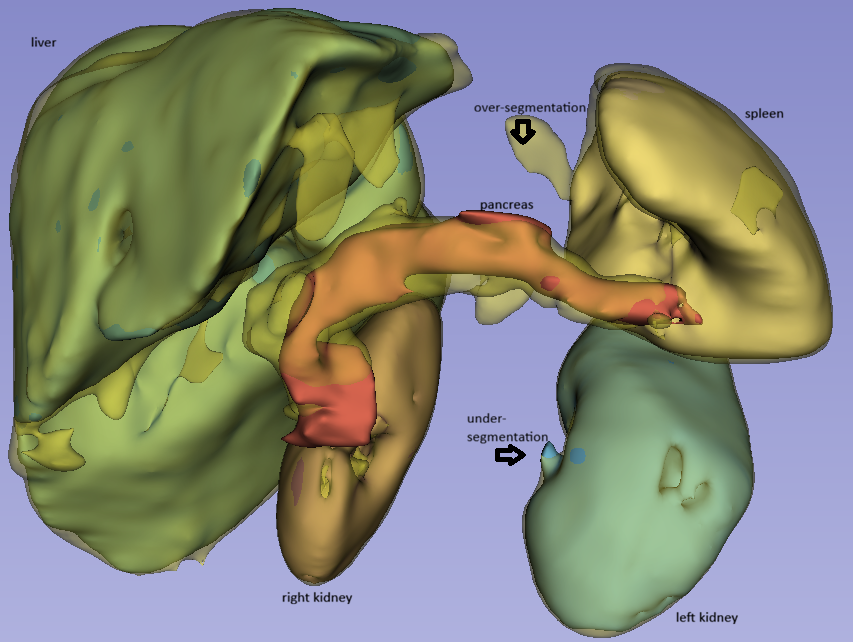}
   \end{tabular}
   \end{center}
   \caption[example]
   { \label{fig:ogr} 
Organ group ground truth left to right: liver=green, spleen=light orange, right kidney=brown, left kidney=blue, pancreas=red. Predicted segmentations all half-transparent yellow, each. Most ground truth colors have a yellowish shimmer around them like a hull. Major error source then is: over-segmentation/fit (prediction locally too big vs. ground truth). Seldomly, the pure groundtruth color pops through, the rarer error then is: under-segmentation/fit (prediction locally too small vs. ground truth). Kidneys are easy, pancreas is most difficult, because of small thin, elongated, snaky shape and low contrast. It is even hard and error prone to manually segment for medical experts.}
\end{figure*}

\textcolor{blue}{
Qualitatively, a total surface rendering from one angle anterior, slightly inferior and left angle is shown (Fig. \ref{fig:ogr}).
For each organ, we show surface renderings of a sample outcome in opposite view angles, so that the total surface coverage gets clear (Figs. \ref{fig:2sub_a}-\ref{fig:sub_b4}).
}

Quantitatively to evaluate the effect of adding a probability map to 2D multi-planar organ segmentation, two different model approaches from section \ref{sec:multi-planar} are trained and compared on a dataset of 80 scans described in section \ref{sec:pre}. The first model (without $SOM$) ensemble uses the scan as single input channel. The second competing model ensemble (with $SOM$) uses two input channels consisting of the scan and the spatial occurrence map. A 5-fold randomized cross-validation with 4:1 splits was used for our study results in Tab. \ref{tab:dice}. For fair comparison, in each cross validation fold, we use the same randomly selected 64 training scans and labels for both model ensembles \textit{with SOM} vs. \textit{without SOM}, and test with the same separate training and test sets (16 scans). To assess the overlap of segmentation outputs vs. ground truth, we use the Dice similarity coefficient (DSC) as a metric. We calculate means and standard deviations, medians and Inter-Quartile-Ranges (IQR) as measures of accuracy and precision.

\section{Results} \label{sec:results} \label{sec:res}
The results for a sample segmentation result and then the quantitative results are shown.

\textcolor{blue}{ 
\subsection{Qualitative results}
In Fig. \ref{fig:ogr} we show a typical organ group, where pancreas as the thinnest elongated reddish structure is most difficult to segment with huge portions of local over- and under-segmentation areas. In the following sub-chapters we show each organ surface completely by two opposite views.
}
\subsubsection{Liver}
\begin{figure*} [!ht]
   \centering
    \begin{subfigure}[t]{\columnwidth}
        \centering
        \includegraphics[width=\linewidth]{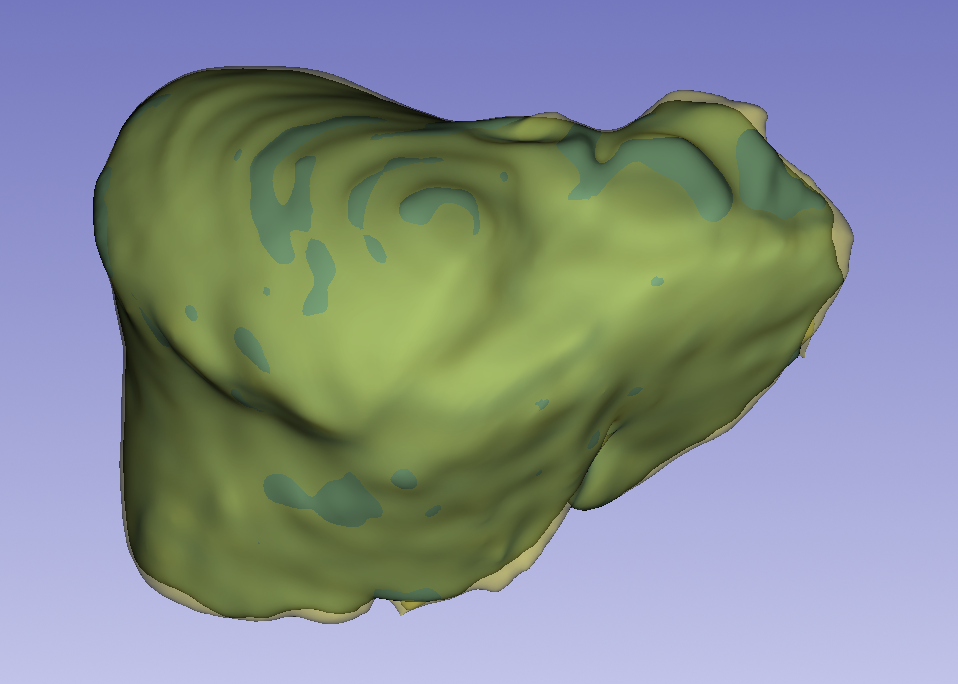}
        \caption{right-superior view}
        \label{fig:2sub_a}
    \end{subfigure}
    \hfill
    \begin{subfigure}[t]{\columnwidth}
        \centering
        \includegraphics[width=\linewidth]{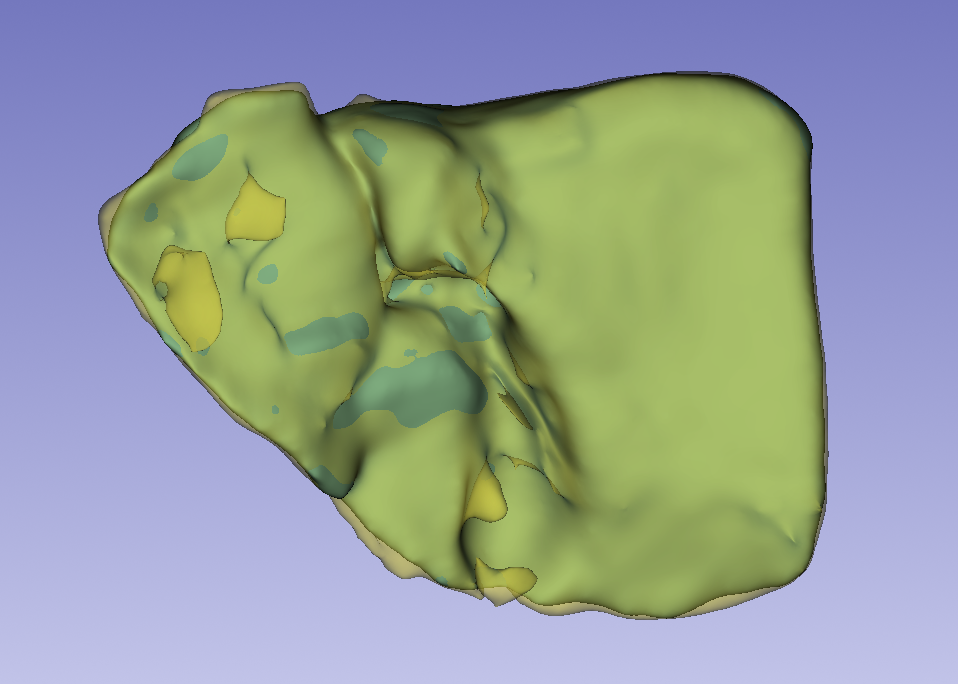}
        \caption{left-inferior view}
        \label{fig:2sub_b}
    \end{subfigure}
\vspace{2mm}
    \caption{Liver=green. Predicted segmentations=half-transparent yellow. (a) Peripheral surface represents well fit. (b) Over-segmentation artifacts show as spiky noses (yellow).}
    \label{fig:main1}
\end{figure*}

The liver peripheral area is fitted well (Fig. \ref{fig:2sub_a}), while the viscerally oriented area shows spikes (Fig. \ref{fig:2sub_b}).

\subsubsection{Spleen}
\begin{figure*} [!ht]
   \centering
    \begin{subfigure}[t]{\columnwidth}
        \centering
        \includegraphics[width=\linewidth]{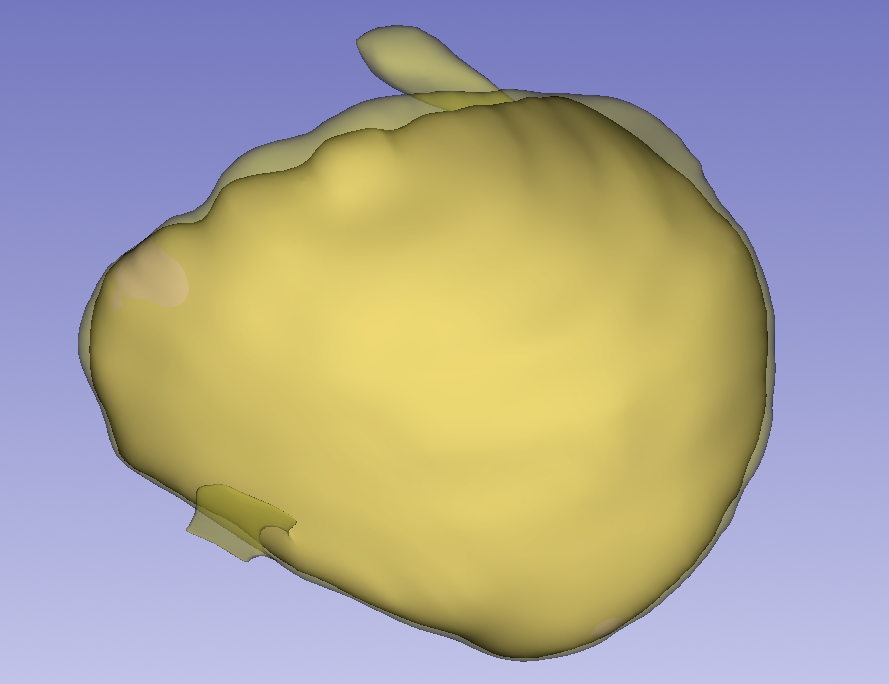}
        \caption{left-superior view}
        \label{fig:sub_a1}
    \end{subfigure}
    \hfill
    \begin{subfigure}[t]{\columnwidth}
        \centering
        \includegraphics[width=\linewidth]{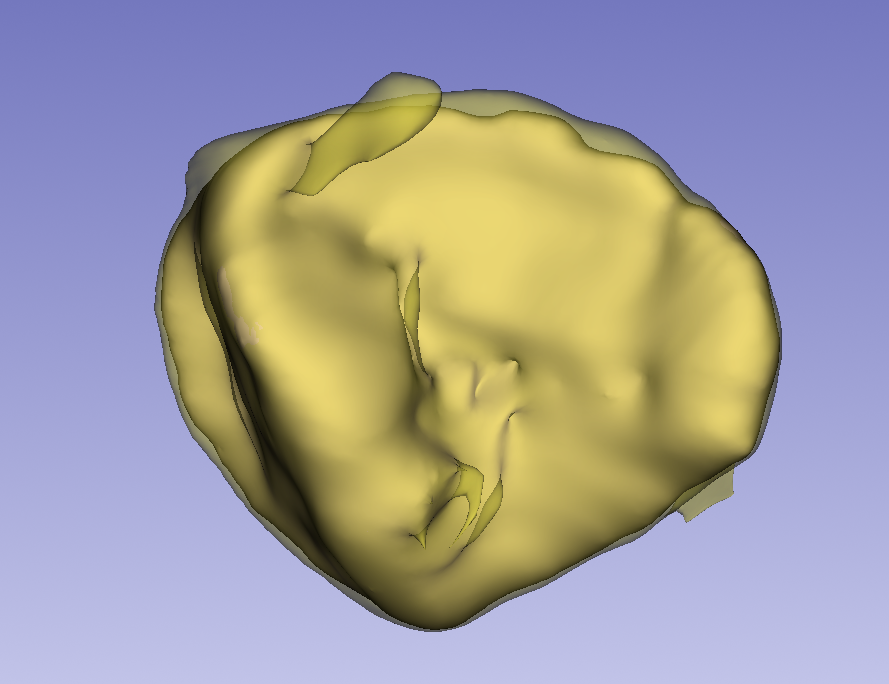}
        \caption{right-inferior view}
        \label{fig:sub_b1}
    \end{subfigure}
\vspace{2mm}
    \caption{Spleen=orange. Predicted segmentation=half-transparent yellow. (a) Peripheral area seems a good fit. (b) Some over-segmentation spikes toward the visceral cavity.}
    \label{fig:main2}
\end{figure*}

The outside segmentation as for the liver looks fairly good (Fig. \ref{fig:sub_a1}). The spleen as well as the liver show some over-segmentation spikes pointing to the visceral cavity of the patient (Fig. \ref{fig:sub_b1}).

\subsubsection{Right kidney}
\begin{figure*} [!ht]
   \centering
    \begin{subfigure}[t]{\columnwidth}
        \centering
        \includegraphics[width=\linewidth]{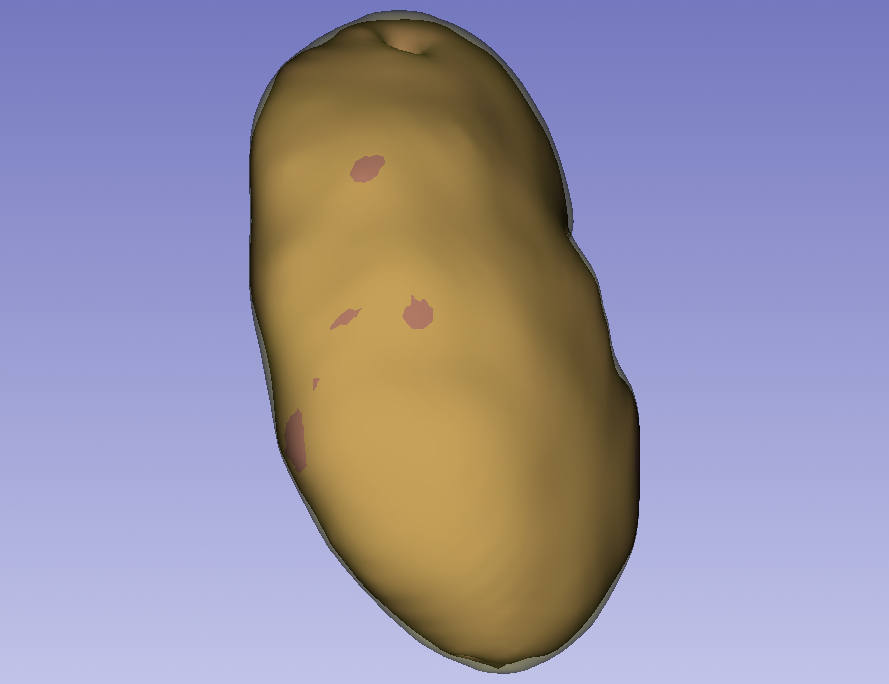}
        \caption{right view}
        \label{fig:sub_a2}
    \end{subfigure}
    \hfill
    \begin{subfigure}[t]{\columnwidth}
        \centering
        \includegraphics[width=\linewidth]{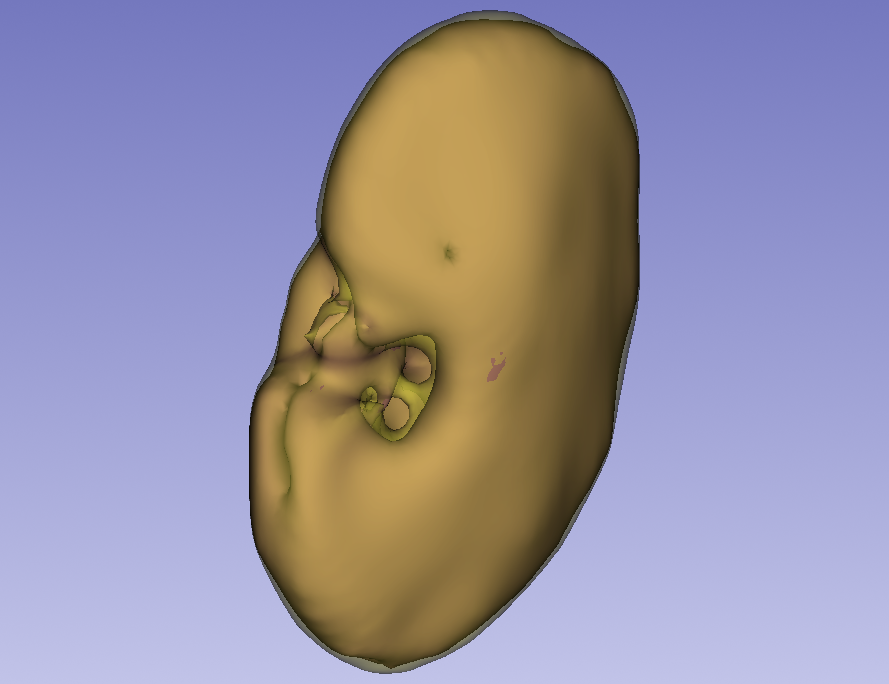}
        \caption{left view}
        \label{fig:sub_b2}
    \end{subfigure}
\vspace{2mm}
    \caption{Right kidney=brown. Predicted segmentations=half-transparent yellow. (a) The kidney is well fit. (b) Same from the visceral direction.}
    \label{fig:main3}
\end{figure*}

The right kidney shows a good segmentation surface approximation (Figs. \ref{fig:sub_a2}, \ref{fig:sub_b2}). As for the others, slight over-segmentation is dominant.

\subsubsection{Left kidney}
\begin{figure*} [!ht]
   \centering
    \begin{subfigure}[t]{\columnwidth}
        \centering
        \includegraphics[width=\linewidth]{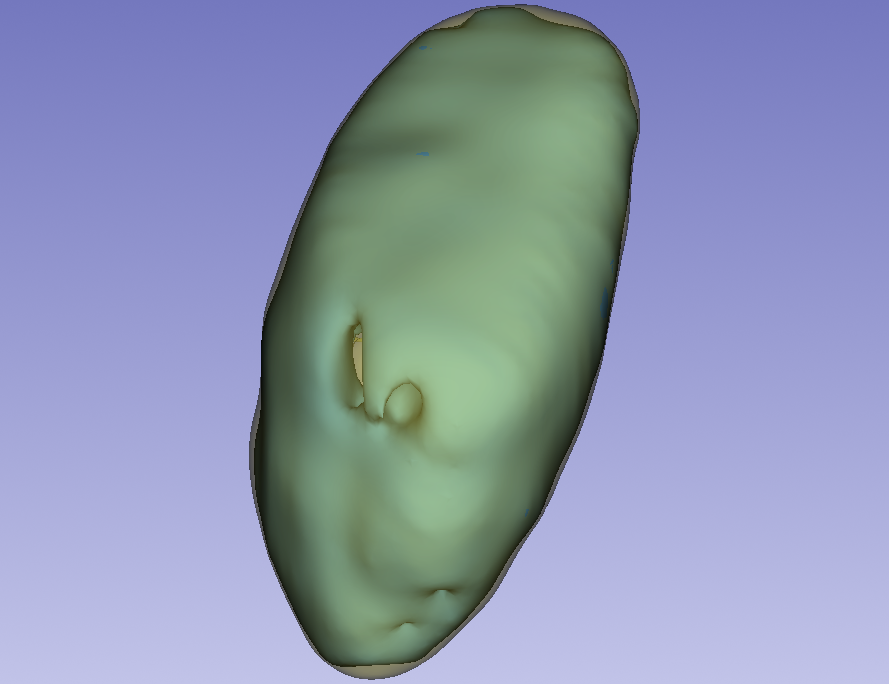}
        \caption{left view}
        \label{fig:sub_a3}
    \end{subfigure}
    \hfill
    \begin{subfigure}[t]{\columnwidth}
        \centering
        \includegraphics[width=\linewidth]{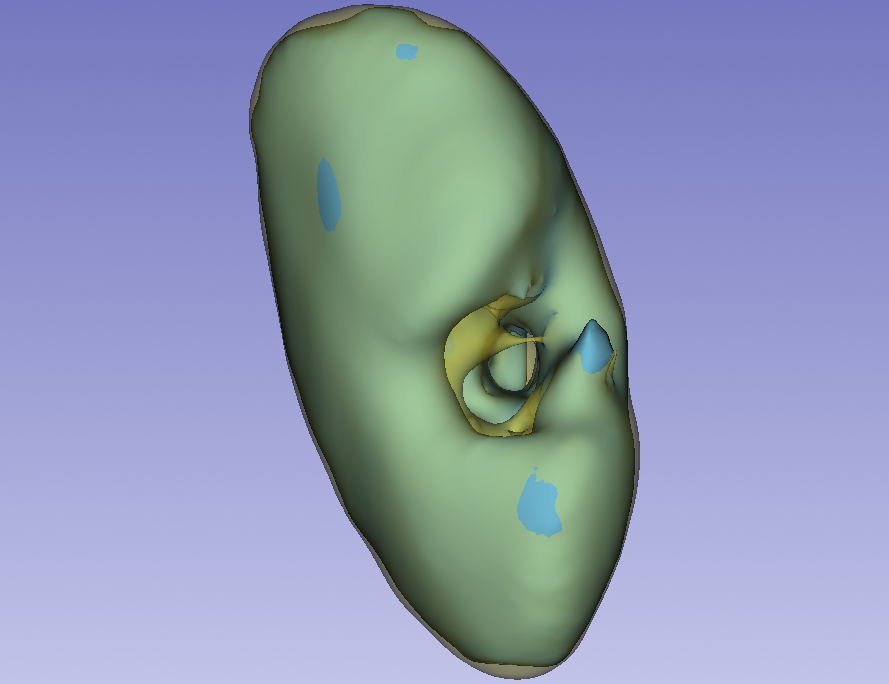}
        \caption{right view}
        \label{fig:sub_b3}
    \end{subfigure}
\vspace{2mm}
    \caption{Left kidney=blue. Predicted segmentations=half-transparent yellow. (a) The kidney fit also reproduces a small hole in the middle, also present in the ground truth data. (b) Good fit from the visceral direction.}
    \label{fig:main4}
\end{figure*}

The left kidney fit also reproduces a patient individual hole present in the patient ground truth data, see Fig. \ref{fig:sub_a3}. The opposite side inside-out of the patient represents again a good fit of this kidney side (Fig. \ref{fig:sub_b3}).

\subsubsection{Pancreas}
\begin{figure*} [!ht]
   \centering
    \begin{subfigure}[t]{\columnwidth}
        \centering
        \includegraphics[width=\linewidth]{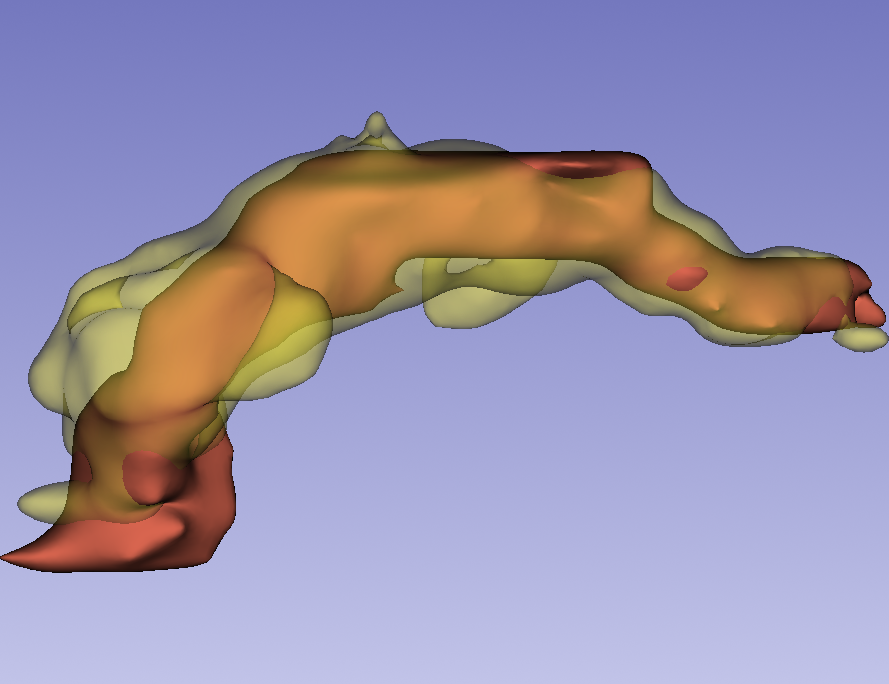}
        \caption{anterior view}
        \label{fig:sub_a4}
    \end{subfigure}
    \hfill
    \begin{subfigure}[t]{\columnwidth}
        \centering
        \includegraphics[width=\linewidth]{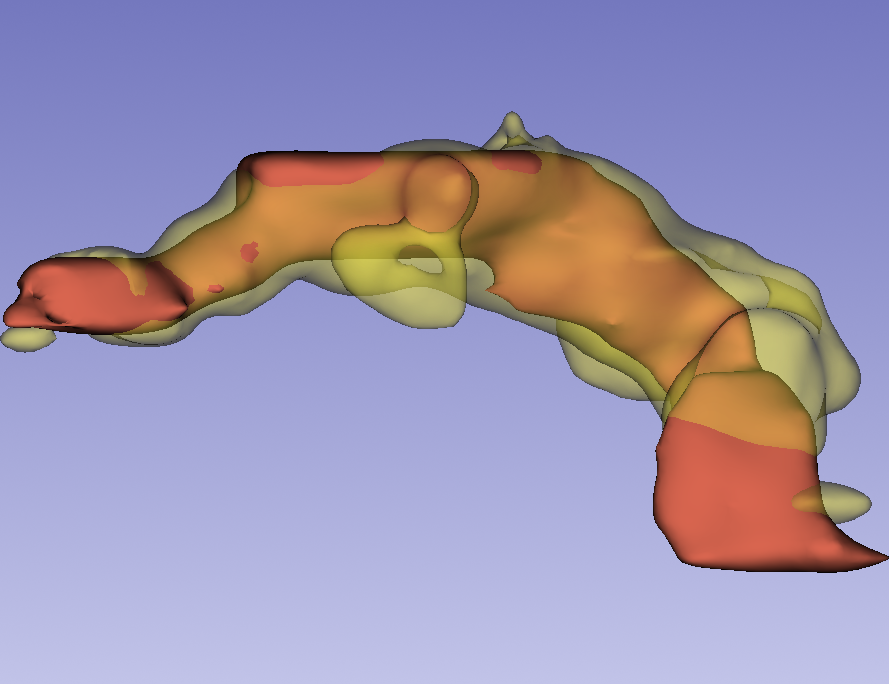}
        \caption{posterior view}
        \label{fig:sub_b4}
    \end{subfigure}
\vspace{2mm}
    \caption{\label{fig:panc} Pancreas=red. Predicted segmentations=half-transparent yellow. (a) The segmentation errors of this purely visceral organ are apparently biggest, also in (b)'s viewing direction.}
    \label{fig:main5}
\end{figure*}

The pancreas, as the hardest to segment object in the ensemble, shows in Figs. \ref{fig:sub_a4} and \ref{fig:sub_b4} the largest segmentation surface fit errors of the 5 organ ensemble.

\begin{table*}[!ht]
\vspace{5mm}
\caption{Mean DSCs with standard deviations (mean$\pm$sd) and median DSCs with Inter-Quartile-Range (IQR) (median$\pm$IQR/2) of the two multi-planar ensemble models trained with and without $SOM$ from 5-fold randomized cross-validation experiments using 4:1 splits of the 80 images into training and test data. \textbf{Bold values} denote organ line maximum, values rounded to 2 decimals.}
\centering
\begin{tabular*}{\textwidth}{l@{\extracolsep\fill}l|ll|ll}
\textbf{DSC} & \multicolumn{2}{|l}{\textbf{mean$\pm$sd}}     & \multicolumn{2}{|l}{\textbf{median$\pm$IQR/2}} \\
\textbf{organ}          & \multicolumn{1}{|l|}{\textit{without $SOM$}} & \textit{with $SOM$} & \multicolumn{1}{|l|}{\textit{without $SOM$}}  & \textit{with $SOM$} \\ \hline
\textbf{liver}          & \multicolumn{1}{|c|}{0.93$\pm$0.03}        & \textbf{0.94$\pm$0.03}     & \multicolumn{1}{|c|}{0.94$\pm$0.02}             & \textbf{0.95$\pm$0.01}         \\ \hline
\textbf{spleen}         & \multicolumn{1}{|c|}{0.92$\pm$0.06}        & \textbf{0.93$\pm$0.04}     & \multicolumn{1}{|c|}{0.93$\pm$0.03}             & \textbf{0.94$\pm$0.02}         \\ \hline
\textbf{right kidney}   & \multicolumn{1}{|c|}{0.90$\pm$0.08}        & \textbf{0.92$\pm$0.05}     & \multicolumn{1}{|c|}{0.92$\pm$0.03}             & \textbf{0.93$\pm$0.02}         \\ \hline
\textbf{left kidney}    & \multicolumn{1}{|c|}{0.91$\pm$0.07}        & \textbf{0.92$\pm$0.05}     & \multicolumn{1}{|c|}{0.93$\pm$0.02}             & \textbf{0.93$\pm$0.03}         \\ \hline
\textbf{pancreas}       & \multicolumn{1}{|c|}{0.67$\pm$0.16}        & \textbf{0.70$\pm$0.13}     & \multicolumn{1}{|c|}{0.71$\pm$0.08}             & \textbf{0.74$\pm$0.07}        
\end{tabular*}
  \label{tab:dice}
\end{table*}

\subsection{Quantitative results}
The evaluation results in Tab. \ref{tab:dice} show, both models achieve mean DSC values above 90\% for all organs except pancreas. For liver, spleen and kidneys, slightly better mean values with an increase between 1 and 3\% result for the 3 fused 2D-U-Net \textit{with SOM}-models. The biggest difference of about plus 3\% occurs when comparing between \textit{without SOM} vs. \textit{with SOM} the last rows' mean and median values in Tab. \ref{tab:dice}. However, pancreas shows larger standard deviation errors, cf. again Fig. \ref{fig:ogr}.

Higher mean values and lower standard deviations across all segmentation results consistently indicate higher accuracy and precision of the method using the $SOM$s in Tab. \ref{tab:dice}. IQRs are also lower in all organs except the left kidney.

Model fit timing is 15 hours for a 5-fold cross-validation on Nvidia A100 GPU (20 GB VRAM used), for experiments \textit{with SOM} and \textit{without SOM}, each. Prediction time is very fast (<1 sec) and hence negligible, which is practically important for model users. That breaks down to a training plus evaluation (1-fold) time of 3 hours for a complete 4-ensemble of 2D-U-Net models (1. axial VOI localization model; 2. axial, 3. coronal and 4. sagittal direction models inside VOI-bounds). Regarding practical application on one patient image, we yield 15 minutes to build one 4-ensemble model from 64 training images and segment one patient (negligible). The number of parameters for each directional 2D-U-Net model is 481,974 (axial VOI-localization) resp. 1,941,046 (inside VOI), totalling to 3 times latter value of finally cumulated 481,974 + 3$\times$1,941,046 = 6,305,112 for a 4-ensemble. \footnote{A corresponding ad-hoc approach of combining 2 3D-U-Net models (1. VOI localization; 2. 3D model inside VOI-bounds) would consume 2$\times$5,645,430 = 11,290,860 parameters.}

\section{Discussion} \label{sec:discussion}
This paper presents a computationally efficient method to locate the upper abdominal body region and perform multi-organ segmentation based on multi-planar 2D slices enhanced by a spatial occurrence map. \textcolor{blue}{\marginpar{3.5, 3.11} The resampling voxel size of 2 mm$^3$ is a compromise between the worlds of general anatomy (1-1.5 mm) and whole-body segmentation or big data studies (2-3 mm). We decided on this balance based on lower GPU-VRAM demands, as our method is also intended to run on consumer-grade GPUs. Moreover, we do not focus on a single organ, but look at a multi-organ group covering longitudinally ca. 30 cm of the adult human body. We acknowledge that this choice may be somewhat unfavorable with respect to partial volume effects and boundary accuracy, but analyzing these factors is beyond the scope of this study.} 
\textcolor{green}{\marginpar{3.1} The comparison to our previous work \cite{0000-2dvs3dunet} shows best results' improvements, but lacks fair comparability as the method pipeline and image data base has changed due to development steps. This also highlights the need for caution when comparing studies too narrowly based on the highest Dice scores alone.}
\textcolor{blue}{\marginpar{2.1, 3.9, 3.12}While this work reports no disruption to advance the field, it proofs a basic principle and shows consistent improvements of this feature $SOM$ in a clear $\Delta$ one factor variation (addition rather than ablation) study design. It is close at hand, additional domain specific $SOMs$ could help other pipelines, too. So, the beneficially sorted out factor $SOM$ could also help other researchers improving their (some structures even better e.g. 98\%) results towards 99\% Dice scores.} \textcolor{blue}{\marginpar{2.2} The one remaining manual design is a choice of first gradient: 1=superior, 0=inferior or the other way around (ascend, decent). Secondly, inverted gradient map and the occurrence maps, which are generated fully automatically from the training data, are combined. We suppose the most positive influence to Dice results to originate from the $SOM$ calculated from organ training shapes. The fuzzy logic inspired directed gradient may be regarded barely significant, which we observed in a side experiment switching the gradient map, and yielding 1\% Dice advantage for liver and spleen using the proposed map from this work. The other map focusing on lower organs (kidneys, pancreas) was better for those by 2-3\% using our pipeline. Thus, this one manual map choice depends really on the application domain, i.e. here targeted organs. For full automatization, just leaving out the manual choice map is a sound way to conclude from here or use a non-directed symmetric gradient map.} \textcolor{blue}{\marginpar{2.4} Similar recent ground works are from 2017 \cite{vaswani2023attentionneed} and caused some interesting follow-up studies on concrete objects in computer vision. However, still the basic 2017 idea is not completely new as computational modeling of visual attention probability maps already started at least in 2001 \cite{itti2001computational} to highlight important image regions. The difference to our architecture line of further developments is the incorporation of assisting maps channel-wise (a-priori) into the input layer of the U-Net, while other works compute hint maps along encoding and decoding paths of the net design. We thus suppose our approach to be beneficially faster, easier to implement and reproduce for other experimenting researchers.} The liver as the biggest and easiest structure achieves the highest mean/median Dice scores accompanied by lowest deviations, see Fig.~\ref{fig:ogr} top-left. Then the spleen follows as the second largest organ, see Fig.~\ref{fig:ogr} top-right. The kidneys consist of round convex shape and come in third, see Fig.~\ref{fig:ogr} low-left and -right. The pancreas as the smallest, thin an elongated structure concludes this list, see Fig. \ref{fig:ogr} middle area. At this point it should be noted that the nature of the Dice score favors large volume structures and segmentation methods favor rounded organs such as the kidneys~\cite{0000-multiOrganDiceAI}. Thus, there is a systematic overall bias in the factors race, gender (females/children: smaller organs) and age (aged: smaller organs) \cite{llopis2026diverse}. \textcolor{blue}{\marginpar{1.1}Therefore, highlighting bias-affected organ volumes of interest using our proposed bounding boxes and SOMs could help alleviate size bias in future work. E.g., by triggering locally adaptive data resampling after the VOI localization phase such that between different groups of humans the same number of voxels results inside a bounding box for an organ or structure}. Due to the sequential order of the training of the individual models, only 6 million U-Net parameters have to be trained vs. more than ten million parameters in an equivalent 3D-U-Net approach. The method performs efficiently and successfully with consumer GPUs (e.g. 16 GB VRAM), too. The evaluation fold models were trained with 64 and tested with 16 CT and label images, so as of now our 2D-U-Net model fusion approach presents a promising and light ensemble architecture already yielding good results. \textcolor{blue}{\marginpar{2.6c}The gradient, spatial occurrence maps and their combination used in this report are of simple and clear design and of course can be further improved. The SOM-design in other domains should be reasonably optimized to specific organs or structures under study. We see the benefit of providing additional input channels being due to more information density, i.e. feature vector information, in relevant areas. In summary, color images (multi-channel RGB) are just simply more discriminative than gray images. Imagine  differentiating red ties vs. blue ties in color vs. gray imagery. Our additional prior maps have to be digested during network training and possibly result in an advantage in the trained weights. The benefit of the U-Nets motivated $SOM$ augmentation principle is shown in this study by consistent mean and median trend improvements in Tab. \ref{tab:dice}.}

\textcolor{blue}{
We are also currently conducting a sub-project on the manual correction of segmentation artifacts caused by over- and under-segmentation, using VR techniques and haptic force feedback \cite{mastmeyer2013ray,fortmeier2013optimized}. In the present results, the liver and spleen exhibit spike-like over-segmentations protruding toward the visceral cavity (Figs. \ref{fig:2sub_b}, \ref{fig:sub_b1}). These artifacts could be corrected by cut-off operations in our visuo-haptic segmentation correction prototype, which is currently under development. The reason for the spikes towards the inner cavity could be due to the difficult and patient-specific neighbor structures, such as stomach and kidneys as well as intestinal structures. With regards to training and prediction, the appearance of these visceral structures is dependent on food and fluid alimentation and not so regular compared to peripheral surfaces. Thus, outer organ surfaces are captured better by the predictions (Figs. \ref{fig:2sub_a}-\ref{fig:sub_a3}, (a)'s). Obviously, the highest challenge in VR correction is also the pancreas, see Fig. \ref{fig:panc}, with it's thin and snake-resembling shape and also being completely surrounded by shape- and intensity-variable visceral structures (Fig. \ref{fig:1sub_a1}, half-filled stomach up-right). Interactive correction could be effectively achieved using specialized pancreas visualization and correction tools.}

\section{Conclusion}\label{sec:conclusion}
We 
positively answer the research question from Section \ref{sec:quest} via showing persistent trend improvements, i.e. by proving a plausible cause (\textit{with SOM}) and effect (higher means, lower deviations) regarding the new U-Net method augmentation by an additive distinct spatial map channels separated from the main image channel. To this aim, we used a statistically sound number of patient scans of 80 with according multi-label expert segmentation masks. For training and evaluation we feature a 5 fold cross-validation study design to account for robustness and reproducibility. In our further future work, we will also look into the effects of more organ-specific $SOM$s, more separate channels (e.g. linear map vs. organ occurence map), architectural variation, i.e. e.g. compare the SOM-approach fair when used inside the 3D-U-Net, larger data bases and liver pathologies. In general, \textcolor{blue}{\marginpar{2.7}our simple yet promising basic principle concept opens a field of future research and questions, and in the practical application field specialization opportunities to different computer vision problems, structures and objects. It was exemplified on a medical imaging example with heuristic map choices.}

\section*{Acknowledgments} \label{sec:ack}
German Research Foundation: DFG MA 6791/1-1.\\
Figure visualizations made with 3D-Slicer \cite{Kikinis2014},\\
and ITK-Snap \cite{py06nimg}.

\section{References} \label{sec:references}

\printbibliography[heading=none]

\end{document}